\documentclass[12pt]{article}
\usepackage{times}
\usepackage{geometry}
\usepackage[utf8]{inputenc}
\usepackage{enumitem,amssymb}
\usepackage{ragged2e}
\usepackage{graphicx}
\usepackage{comment}
\usepackage{multicol}
\usepackage[usenames]{xcolor} 
\definecolor{xlinkcolor}{cmyk}{1,1,0,0}
\usepackage{url}
\usepackage[
 colorlinks=true,    
 linkcolor=xlinkcolor,     
 citecolor=xlinkcolor,     
 filecolor=xlinkcolor,     
 urlcolor=xlinkcolor,      
 final=true
]{hyperref}
\usepackage[super,sort&compress,comma]{natbib}
\usepackage{enumitem}
\setenumerate{itemsep=0mm}

\begin{document}


\title{Using TeV Cosmic Rays to probe the Heliosphere's Boundary with the Local Interstellar Medium}

\date{}
\maketitle
\vspace{-1.cm}

\begin{center}
Paolo Desiati$^1$,
Juan Carlos D\'iaz V\'elez$^{1,2}$,
Gwenael Giacinti$^3$,
Francesco Longo$^4$,
Elena Orlando$^{4,5}$,
Nikolai Pogorelov$^6$,
Ming Zhang$^7$
\end{center}
{\footnotesize
$^1$ Wisconsin IceCube Particle Astrophysics Center (WIPAC), University of Wisconsin, Madison, U.S.A.\\
$^2$ Universidad de Guadalajara, Guadalajara, Jal., M\'exico\\
$^3$ Tsung-Dao Lee Institute, Shanghai Jiao Tong University, Shanghai, China\\
$^4$ University of Trieste \& National Institute for Nuclear Physics (INFN), Italy\\
$^5$ Kavli Institute for Particle Astrophysics and Cosmology (KIPAC) \& Hansen Experimental Physics Laboratory, Stanford University, U.S.A.\\
$^6$ University of Alabama in Huntsville, U.S.A.\\
$^7$ Florida Institute of Technology, U.S.A.\\
}

\noindent {\normalsize \bf Abstract:}
%
%
{\small
The heliosphere is the magnetic structure formed by the Sun's atmosphere extending into the local interstellar medium (ISM). The heliopause, the boundary separating the heliosphere from the ISM, is a still largely unexplored region of space. Even though the Voyager spacecraft officially entered the local ISM in 2012 (V1) and 2018 (V2) and are delivering data on the outer space environment, they are just two points piercing a vast region of space at specific times. The heliospheric boundary regulates the penetration of MeV-GeV galactic cosmic rays (CR) into the inner heliosphere, where the solar system is located. Interstellar keV neutral atoms are crucial to the outer heliosphere since they can penetrate unperturbed and transfer energy into the solar wind. Missions such as NASA's Interstellar Boundary EXplorer (IBEX) and Cassini are designed to detect neutral atoms and monitor charge exchange processes at the heliospheric boundary. The heliosphere does not modulate the intensity of TeV CR particles coming from the ISM, but it does influence their arrival direction distribution. Ground-based CR observatories have provided statistically accurate maps of CR anisotropy as a function of energy over the last couple of decades. Combining such observations to produce all-sky coverage makes it possible to investigate the impact that the heliosphere has on TeV CR particles. We can numerically calculate the pristine TeV CR pitch angle distribution in the local ISM using state-of-the-art heliosphere models. Only with the heliospheric influence subtracted is it possible to use TeV CR observations to infer propagation properties and the characteristics of magnetic turbulence in the ISM. Numerical calculations of CR particle trajectories through heliospheric models, therefore, provide a complementary tool to probe into the global properties of the boundary region, such as its size, length, and the scale of the local interstellar magnetic field draping around the heliosphere. A program boosting heliospheric modeling with emphasis on the boundary region, and promoting combined CR experimental data analyses from multiple ground-based experiments, will benefit CR astrophysics and, in reverse, will provide additional data and complementary tools to explore the interaction between the heliosphere and the local ISM.
}

\clearpage

\setcounter{page}{1}
\section{Introduction}
\label{sec:intro}
%
%
Below a few tens GeV, CR particles are directly influenced by inner heliospheric dynamics driven by the Sun. Such effects are observed in terms of flux modulation following the solar cycle~\cite{pamela, ams02}.
The most extensive studies of CR modulations are done by monitoring the secondary neutrons on the ground generated by primary CR-induced cascades with Earth's atmosphere~\cite{neutron}. Neutron monitoring is an effective method to study the Sun's eleven-year magnetic field polarity inversions and the inner heliosphere's variability that follows. Moreover, direct CR observations with high-altitude and space-borne instrumentation~\cite{crdirect}, provide additional information to investigate CR propagation~\cite{strong1998, blasi2017} and inter-planetary magnetic turbulence over the course of solar cycles~\cite{bon2020}. 
Finally, data collected by Voyager in the outer heliosphere offer additional boundary conditions for new heliospheric models and connecting CR flux in the inner and outer heliosphere's regions~\cite{voyager}. However, Voyager data hardly provide sufficient information able to constrain the whole heliospheric boundary region with the ISM. Full-sky observations with IBEX~\cite{ibexmap} and the future Interstellar Mapping and Acceleration Probe~\cite{imap} (IMAP) are designed to deliver crucial data feeding into state-of-the-art heliospheric modeling. Such models must be able to coherently describe the dynamics of keV energetic neutral atoms and TeV CRs.

CR particles at TeV-scale and above are insensitive to the inner heliospheric modulations, as their gyroradius is larger than the termination shock. At these energies, however, they are expected to be affected by the large scale structure and the outer region of the heliosphere. Rather than a solar cycle modulation, the impact of the heliosphere on TeV CRs primarily manifests as a {\it lensing effect} modifying their arrival direction distribution shaped by the ISM. Understanding how the heliosphere modifies the TeV CR distribution is crucial, if we want to investigate propagation in the ISM, interstellar turbulence and, ultimately pinpoint nearby source contribution to the observed flux.

Modern operating and future ground-based CR observatories are designed to collect a large number of shower events up to ultra-high energies, with unprecedented statistical accuracy and increasing experimental precision~\cite{crindirect}. The energy, mass, and arrival direction distribution of high energy CRs provide observational hints of the complex environment through which CRs propagate from their sources to Earth: the source turbulent environment, the injection into the ISM and propagation through interstellar turbulence, and the effects of coherent magnetic structures, such as the Local Bubble and the heliosphere.
If the {\it heliosphere's foreground} is not taken into account, astrophysical interpretations of high energy CR observations result biased. In other words, if we want to extract astrophysical information from CR observations, we must account for the distorting effects imprinted by the heliosphere.

Similarly to how GeV CRs provide us with key information regarding the inner heliosphere, the observed TeV CRs distributions are shaped by galactic propagation, but also by the properties of the outer heliosphere and the local ISM, making them complementary to keV energetic neutral atoms. The goal is to use TeV CR data to extend models driven by inner heliospheric properties into the outer boundary region with the ISM. Emphasising large scale heliospheric modeling is therefore the new frontier to gain augmented astrophysical information on energetic CRs. On the other hand, heliospheric models aimed to improve our understanding of the boundary region with the local ISM, benefits from using high-volume and high-quality TeV CR observations.

\section{Context}
\label{sec:context}
The primary context to consider consists in understanding how the heliosphere influences high energy CR distributions as observed on Earth. A simple dimensional consideration provides a useful hint: 10 TeV protons have a gyroradius of about 500-800 AU in a 3-5 $\mu$G local interstellar magnetic field. This is about the transverse size of the heliosphere, which implies that some sort of {\it spacial resonance} must occur when CR particles pass through the heliosphere~\citep{drury2008, lazarian_desiati_2010, desiati_lazarian_2012, desiati_lazarian_2013, drury2013, schwadron_2014, zhang_2014, lb_2016, lb_2017, lb_2019, icrc2019, zhang_2020}. CR pitch angle distribution in the ISM is modified by the presence of the heliosphere when particles are collected on Earth. Understanding how this modification occurs requires a clear comprehension of the global heliosphere's structure and unbiased high-quality experimental CR data.

Another context is the possibility to study interstellar turbulence using high energy CR anisotropy distributions, in addition to energy spectra and mass composition. Once the heliosphere's influence is subtracted from the observed CR distributions, the {\it pristine} pitch angle distribution in the ISM can be used to investigate the properties of interstellar magnetic turbulence~\cite{giacinti_sigl_2012, giacinti_kirk_2017, giacinti_2021, ahlers_2014, mertsch2015, mertsch_ahlers_2015, ahlers_mertsch_2016, mertsch_2019, genolini_2021, kuhlen_2022}. With a better understanding of particles' diffusion in the ISM, we can constrain astrophysical propagation models that can potentially provide hints of recent local sources of the observed CRs~\cite{schwadron_2014, ahlers_2016}. It is possible that the observed CR distributions are specific to our local environment and not representative of a galactic-scale phenomenology.

Moreover, a heliospheric study context is possible by comparing state-of-the art modeling with combined high-quality ground-based experimental CR observations~\cite{zhang_2014, icrc2019, zhang_2020}. Numerical particle trajectory integration calculations aimed to predict the CR distributions on Earth given their interstellar pitch angle distribution, proves to be an effective way to probe the large scale properties of the heliosphere (such as, e.g. size, length, and how interstellar magnetic field lines drape around it). By providing new constrains on the global heliosphere's characteristics, we can improve the models and gain additional information on how solar wind interacts with the local ISM. This is not about improving our knowledge of the inner heliospheric mechanisms that lead to space weather, but it's about learning how the heliosphere interfaces with the ISM and how far its influence goes.

\section{Methods}
\label{sec:methods}
The study of the outer heliosphere relies on complex state-of-the-art numerical modeling aimed to reproduce in-situ and remote observations, with the goal of identifying the heliospheric underlying physical processes (see sec.~\ref{ssec:lism}). The investigation of how TeV CRs are influenced by the heliosphere, provides an additional means to validate and extend the models developed against {\it low energy} observations.

High energy CR data collected by modern ground-based experiments are suitable for this kind of study because of their high statistical accuracy and improved understanding of the experimental systematics uncertainties. All-sky maps of 1-100 TeV CR arrival direction distributions as a function of energy and, possibly CR particle mass, are the ideal tools to investigate the outer heliosphere (or very local ISM) as complementary to the already available in-situ and remote observations. This is achieved with combining CR data collected by different experiments located in the northern and southern hemispheres (see sec.~\ref{ssec:cra}).

Test particle numerical trajectories calculations in the heliospheric magnetic field are performed to produce CR data for comparisons with experimental ground-based observations, with varying parameters of the heliospheric models. A Liouville mapping technique, which takes into account the detailed heliospheric magnetic field structure, is employed (see sec.~\ref{ssec:heliocr}). With such a method, it is possible to derive the density gradient and pitch angle distribution of TeV CRs in the local ISM while accounting for particle trajectory chaotic behavior~\citep{lb_2019}, and the residual experimental systematic biases~\citep{icrc2019} affecting the CR anisotropy sky maps.

The development of heliospheric models and detailed simulation of CR particle trajectories require intensive use of computing resources and in order to carry out many similar calculations that share access to fast memory. NSF-sponsored computing centers possess with high-memory CPU and GPU can provide the needed power to conduct the required numerical integration for our CR trajectory analysis.


\subsection{Local ISM in the presence of the heliosphere}
\label{ssec:lism}
%
The possibility of unfolding the effects of the heliosphere from the observed CR arrival direction distribution constitutes a new drive to develop novel detailed heliospheric models with the emphasis on the solar wind-ISM boundary. Recent modeling involves adaptive mesh refinement numerical integration of Magneto Hydro Dynamic (MHD) equations for plasma coupled with multi-fluid or kinetic transport for neutral atoms~\citep{Pogo14,Pogo21,Federico21}. This model, which accounts for all plasma/magnetic field and neutral gas interactions, was originally developed to make predictions for~\emph{Voyager} heliospheric mission \citep{Stone05,Stone08,Stone13} and interpret IBEX observations~\citep{McComas09,McComas20,Jacob10,Erik16}. Now both \emph{Voyager} 1 and 2 are in the local ISM, making in-situ measurements of the local interstellar magnetic field and plasma properties. Observations just outside the heliosphere \citep{Burlaga22,Gurnett21} can greatly constrain the local ISM parameters. The model has been validated against numerous in-situ and remote observations, e.g., (\emph{SOHO} Ly$\alpha$ back-scattered emission, Ly$\alpha$ absorption profiles in directions towards nearby stars, \textit{New Horizons} observations in the distant solar wind, in-situ measurements in the solar wind and local ISM from \textit{Voyagers}, etc.\citep{Kim16, Kim17, Pogo17a, Pogo17b,Pogo21}. Although we do not expect the local ISM conditions to change within decades of CR measurements, the most recent models take into account solar cycle effects with the input of remote measurements of the photospheric magnetic field and initiate coronal mass ejections using multi-viewpoint observations \citep{Talwinder18, Talwinder19,Talwinder20,Talwinder22}. In this way, it is possible to investigate the potential time-dependence of CR anisotropy. Since TeV CR are sensitive to the transverse size of the heliosphere, in particular to the draping of the heliopause by the interstellar magnetic field lines, the study of CR flux in the local ISM may provide important information about the physical processes occurring in the very local ISM -- the part of the local ISM affected by the presence of the heliosphere \citep{Pogo15,Pogo17a}.

\subsection{Cosmic ray anisotropy}
\label{ssec:cra}
Several observations from large ground-based experiments have provided evidence of a faint (up to order $10^{-3}$) but significant anisotropy of the CR flux at energies above 100 GeV. This is especially true for the TeV-PeV energy range, which is covered by multiple experiments both in the northern and southern hemispheres~\citep{nagashima_1998, hall_1999, amenomori_2005, amenomori_2006, amenomori_2007, guillian_2007, abdo_2008, abdo_2009, aglietta_2009, zhang_2009, munakata_2010, amenomori_2011, dejong_2011, shuwang_2011, bartoli_2013, abeysekara_2014, bartoli_2015, amenomori_2017, bartoli_2018, abeysekara_2018, abbasi_2010, abbasi_2011, abbasi_2012, aartsen_2013, aartsen_2016, lhaaso_2021}. With increasing data volume and substantial improvements in analysis techniques, these observations reveal the CR anisotropy as a function of energy and angular scale with statistical accuracy in relative intensity below $10^{-5}$. However, the limited field of view of any individual ground-based experiment prevents us from capturing the anisotropy features at large angular scale. This limitation cause correlations between the spherical harmonic components used to study the angular structure of the anisotropy~\citep{jcdv_2017}. Combining observations from ground-based experiments located in different hemispheres, we can eliminate such correlations. All-sky arrival direction distribution maps make it possible to investigate CR properties in the context of propagation through the ISM magnetic turbulence~\cite{giacinti_sigl_2012, giacinti_kirk_2017, giacinti_2021, ahlers_2014, mertsch2015, mertsch_ahlers_2015, ahlers_mertsch_2016, mertsch_2019, genolini_2021, kuhlen_2022}.
The HAWC gamma-ray and the IceCube neutrino observatories (located in Mexico and Antarctica, respectively) have compiled the first combined sky map of 10 TeV CR anisotropy~\citep{jcdv_2017}. Undergoing analyses will provide sky maps at higher energies, as well. In the future, using new experiments currently in construction, like LHAASO~\citep{disciascio_2016}, or in design phase, like SWGO~\citep{swgo_whitepaper, swgo_astro2020, swgo_paper} we will be able to construct statistically accurate combined CR distribution sky maps up to PeV energy scale as a function of the primary particle mass. Combining data from different experiments resolves the biases derived from the individual observations' limited field of view on the large-scale angular features~\cite{icrc2019}. All-sky CR arrival direction distribution maps yields a powerful tool to explore the origin of the observed CR anisotropy. In particular, they make it possible to investigate how the outer heliosphere influences CR particle trajectories coming from the ISM~\cite{lazarian_desiati_2010, desiati_lazarian_2012, desiati_lazarian_2013, schwadron_2014, zhang_2014, lb_2016, lb_2017, lb_2019, zhang_2020}.

The observations show that the CR anisotropy changes its angular structure as a function of energy. The global relative excess appears to shift direction across the sky between 100 TeV and 300 TeV, with its shape changing accordingly while staying relatively stable at lower and higher energies. The change in direction cannot be described as a rotation but rather as a transition between two anisotropies, one dominating below 100 TeV and one above 300 TeV. This may be a transition between two sources contributing to the observed cosmic rays, most likely local, or from a significant magnetic structure change at larger distance in our local ISM. The observed full-sky angular power spectrum, devoid of biases from the limited field of view, provides a novel tool to probe into CR propagation in the ISM~\cite{giacinti_sigl_2012, giacinti_kirk_2017, giacinti_2021, ahlers_2014, mertsch2015, mertsch_ahlers_2015, ahlers_mertsch_2016, mertsch_2019, genolini_2021, kuhlen_2022}. Heliospheric influence on the CR distribution below 100 TeV, however, severely limits astrophysical interpretations. The outer heliosphere modulations on the CR arrival direction distribution affect the CR pitch angle distribution concealing their astrophysical properties. 
CR anisotropy observations, in fact, show correlations with the heliosphere and its ISM outer region~\citep{jcdv_2017, icrc2019}. For instance, the flux enhancement referred to as {\it region A}, lies along the heliosphere's tail direction, and in association with the location of the $B$-$V$ plane (the plane formed by the interstellar flow velocity and magnetic field directions deep in the LISM). 

Only recovering the pristine CR distribution in the ISM can we investigate their propagation properties. And this is possible using state-of-the-art models of the heliosphere that best reproduces our TeV observations. Ultimately, this will help us reveal the physics of CR diffusion in the ISM and the turbulence affecting it. Future large-area gamma-ray observatories such as SWGO, are designed to boost the sensitivity in the search for astrophysical sources, and the investigation of CR properties from TeV to PeV. The ability to resolve the primary CR particle mass is essential to understand their origin. Preliminary studies show that SWGO will be able to provide a deeper understanding of the CR anisotropy~\cite{swgo_icrc_2021}.

\subsection{The heliospheric boundary and its effects on high energy CRs}
\label{ssec:heliocr}
Heliospheric magnetic and electric fields significantly influence the arrival direction of TeV CRs observed on Earth. The deviation of arrival direction distorts CR anisotropy map that one would see in the pristine local ISM. To understand the properties and physics of CR propagation in the local ISM using observations of TeV CR anisotropy, we must first remove the foreground heliospheric effects hidden in the observations. Such a task is possible with a mapping technique based on the Liouville theorem, which states that particle distribution function is conserved along any particle trajectory~\cite{zhang_2014, zhang_2020}. Assuming that the total electromagnetic force on TeV CRs can be determined from a heliosphere model, we can compute the trajectory of particles traversing through the heliosphere and map the CR distribution function in the pristine local ISM to anisotropy seen on Earth. In this way, we can determine if a distribution of CRs in the local ISM is consistent with anisotropy measurements. The method was first applied to observations of ~4 TeV CR anisotropy obtained by the Tibet Air Shower (AS) array~\cite{zhang_2020}. It was found that 4 TeV CR anisotropy in the local ISM is dominated by a dipole aligned with the local ISM magnetic field. The dipole indicates a strong flow of CRs along the interstellar magnetic field into the northern Galactic halo at an equivalent speed of ~1500 km/s. The heliosphere distorts the dipole anisotropy, generating mid-scale spherical harmonics contributions with $\ell \sim$ 3 - 11. Beyond that, a weak anisotropy is present in higher order harmonics, probably a residual of magnetic turbulence in the local ISM. The power spectrum can provide critical information about the large-scale turbulence of the interstellar magnetic field. The mapping of TeV CRs anisotropy leads to a determination of CR pitch angle distribution, which could be used to infer the properties of particle scattering by large-scale magnetic field turbulence in the local ISM. The inferred CR distribution from Tibet AS experiment at 4 TeV suggests a density gradient of CRs pointing towards the super bubble approximately in the direction of the Vela supernova remnant, which is consistent with Ref.~\cite{schwadron_2014, ahlers_2016}. If this is confirmed, it is a piece of evidence for a significant contribution of CRs from a relatively local source in the Galaxy.

The above results show that anisotropy observations contain valuable information about CR properties and origin that are complementary to traditional studies using spectral and composition measurements. Our analysis uses a data set from a single ground-based experiment. The Tibet AS investigation only covers declination from -10$^{\circ}$ to 60$^{\circ}$. The limited number of particle counts only permits accurate determination of CR anisotropy at small angular scales and energies around a few TeV. To exploit the full potential of this method, all-sky maps from different ground-based experiments at different energies are necessary.
Analysis of CR anisotropy promises the determination of the pristine local ISM CR properties at all these energies. Information about the energy dependence of CR flow, pitch angle distribution, and density gradient can tell a lot about the question of CR origin.

Reproduction of CR anisotropy from the Liouville mapping technique requires an accurate model of the heliospheric magnetic field and plasma distribution. The model is sensitive to the input magnetic field, neutral atom density, and ionization ratio in the local ISM, as well as how the sun-local ISM plasma interaction is treated. From our previous model calculation, TeV CR anisotropy should also weakly depend on solar cycles. Information contained in CR anisotropy can become a test-bed to determine the quality of heliosphere models. For example, TeV CR anisotropy is a key test to confirm if the heliosphere is a {\it comet-like} structure or {\it croissant-shaped}. Answers to questions about the shape, location, and plasma properties of heliospheric bow-wave may also be found in TeV CR anisotropy measurements.

\section{CR anisotropy and magnetic field turbulence in the ISM}
\label{sec:turb}
The pristine CR pitch angle distribution in the local ISM depends on the properties of the local interstellar turbulence within about a CR mean free path from Earth~\cite{giacinti_2021}. The gyroradius of TeV-PeV CR protons is about 10$^{-4}\,-$ 10$^{-1}$ parsec (pc), which is substantially smaller than the typical coherence length of the interstellar turbulence $l_c\sim 1\,-\,10$ pc~\cite{haverkorn_2008}. Particles are therefore expected to preferentially diffuse along the magnetic field lines. The experimental evidence that the observed CR anisotropy is generally ordered along the local interstellar magnetic field~\cite{schwadron_2014, ahlers_2016, jcdv_2017} corroborates this expectation.

Observations show that CR anisotropy is not a pure dipole~\cite{aartsen_2016, abeysekara_2018, jcdv_2017}, which would be expected only in case pitch angle scattering of CR particles in ISM turbulence is isotropic~\cite{giacinti_kirk_2017}. In particular, the distribution above 100 TeV hints at a flattening of the CR intensity in directions perpendicular to local magnetic field lines. Such hints of anisotropic pitch angle diffusion in the ISM make it possible to test MHD turbulence modes, like, e.g., the relative contribution of fast magnetosonic mode turbulence with an isotropic power spectrum~\cite{cho_lazarian_2002a} and so-called Goldreich-Sridhar anisotropic turbulence~\cite{cho_lazarian_2002b}.

The investigation of long-baseline pitch angle diffusion in the ISM requires statistically accurate combined all-sky observations of CR anisotropy as a function of energy and mass, which is the main goal of present and future ground-based experiments, such as IceCube, HAWC, LHAASO and SWGO. In addition, the understanding and unfolding of the heliospheric influence on the CR distribution below 100 TeV will make it possible to study the role of local interstellar turbulence at lower energy, as well.

\section{The Sun as a steady-state gamma-ray source}
\label{sec:gammasun}
In addition to the science case explained above, SWGO would allow observations of gamma rays from the Sun. In fact, Galactic CRs propagate in the heliosphere and interact with the solar surface and its surroundings. This generates non-thermal steady emission from the quiet Sun that can extend up to hundreds of GeV.
Its steady emission is daily detected by the Fermi LAT telescope \citep{Abdo2011, Barbiellini, Bartoli, Linden, Linden2020, Ng, Tang} and it was previously detected for the first time by \citet{Orlando2008} in the EGRET data, who reported the detection of two components of the emission with a different spatial and spectral shape. The HAWC observatory recently reported a gamma-ray signal from the quiet Sun around 1 TeV and most of it correlates with the solar minimum data~\cite{hawc-quiet-sun}. These are the disk component produced by CR hadrons interacting with the solar surface \citep{Seckel91, Thompson, Mazziotta, Li, Becker, Guti, Hudson20, Nibl, Zhou} and the spatially extended Inverse-Compton (IC) component produced by CR electrons and positrons on the solar photons \citep{Orlando2006, Stellarics}.
The flux of both gamma-ray components is expected to change over the solar cycle due to the modulation of the CRs in the heliosphere, and it is expected to be anti-correlated with the solar activity. The two emission components have been more clearly disentangled in the Fermi LAT data \citep{Abdo2011}. However, present models are not able to totally describe the observed emission \citep{Abdo2011} and some unexpected features showed up in the data.
An example is a GeV deep and a harder  excess  of  GeV  gamma-ray  flux  that  seems to anti-correlate with the solar  activity,  especially  at  the  highest  energies  of  Fermi-LAT \citep[e.g.][]{Tang}.
Gamma-ray observations of the Sun at GeV and TeV energies with SWGO would improve  our  understanding  of the Sun and its environment, the solar activity, and would allow for searches of new physics. These studies can complement the studies of CRs presented above.  

\section*{Summary and outlook}
\label{sec:summary}
The current recommendation to better understand the outer heliosphere and its interaction with the local ISM, relies on NASA's solar-terrestrial probes, as illustrated by the reference mission IMAP in coordination with the Voyager missions~\cite{decadal_survey}. The structure of the outer heliosphere is important to us because it regulates the CR penetration into near-Earth space and it has a direct role in reducing space radiation and making life possible in our solar system. The serendipitous and surprising discovery of an unpredicted narrow {\it ribbon} of keV neutral atom emission from the outer heliosphere, highlighted that this is an active region of space influencing the inner heliosphere~\cite{McComas09}. In addition, the ribbon's ordering with the local interstellar magnetic field, showed its importance in shaping the outer heliosphere. Finally, the observed time modulations of the ribbon, demonstrate that the heliosphere-ISM interaction is more dynamic than expected.

The emission of neutral atoms from the outer heliosphere provides remote access to the region of space just outside the heliopause where the local ISM magnetic field drapes around the heliosphere. In other words, IBEX/IMAP are sensitive to the properties aligned with the distorted local ISM magnetic field, consistently with the Voyager spacecraft that sense the magnetic field directly influenced by the outer heliosheath. TeV CRs, on the other hand, are sensitive to the local ISM environment within thousands of Astronomical Units (a distance that would take centuries for Voyager and future interstellar probes to reach). High-energy CR particles provide a novel remote sensing of the region of space where the local interstellar magnetic field drapes around the heliosphere. The study of how TeV CRs are influenced by the heliosphere as shaped by the observed keV neutral atoms, gives us the key to unfold their distribution in the ISM.

Current ground-based CR experiments, and new generations of observatories such as SWGO, are essential element of this program for the construction of all-sky map of CR arrival direction distribution in the energy range from TeV to PeV. The availability of such massive and sophisticated instrumentation makes it possible to already use the data collected to perform our novel remote sensing study. The goal is to use IceCube and HAWC observatories in concert and ensure that proper models of the heliosphere are developed to study CR particles propagation in the ISM. Future experiments such as SWGO in the southern hemisphere and LHAASO in the northern hemisphere, will guarantee the availability of high quality CR data to perform this study in the next decade~\cite{rankin_2022}.


\newcommand{\mnras} {MNRAS}
\newcommand{\apjl}{ApJL}
\newcommand{\apj}{ApJ}
\newcommand{\ssr}{Space Sci. Rev.}
\newcommand{\apjs}{ApJS}
\newcommand{\prd}{PhRvD}
\newcommand{\aap}{A\&A}
\newcommand{\nat}{Nature}
\newcommand{\jcap}{JCAP}
\newcommand{\aaps}{A\&AS}
\newcommand{\prl}{PRL}
\newcommand{\baas}{Bulletin of the American Astronomical Society}
\newcommand{\jgr}{Journal of Geophysical Research}
\newcommand{\apss}{Ap\&SS}
\clearpage

\end{document}